A (T) Correction for Multicomponent Coupled-Cluster Theory

Dylan Fowler, Kurt R. Brorsen*

Department of Chemistry, University of Missouri, Columbia, Missouri 65203, USA

**Abstract**

The (T) and [T] perturbative corrections are derived for multicomponent coupled-cluster theory with single and double excitations (CCSD). Benchmarking shows that multicomponent CCSD methods that include the perturbative corrections are more accurate than multicomponent CCSD for the calculation of proton affinities and absolute energies. An approximation is introduced that includes only (T) or [T] contributions from mixed electron-nuclear excitations, which significantly reduces computational effort with only small changes in protonic properties.

Coupled-cluster (CC) theory with single and double excitations and perturbative triples (CCSD(T))[1] is commonly referred to as the "gold standard" of quantum chemistry due to its high accuracy and acceptable $N^7$ computational scaling with respect to system size. Standard or single-component formulations of CC theory[2] treat all nuclei classically by invoking the Born-Oppenheimer approximation, but there are also multicomponent[3] CC methods[4-9] that allow select nuclei to be treated quantum mechanically and include nuclear quantum effects (NQEs) in the calculations. However, almost all previous multicomponent CC methods have been limited to the inclusion of single and double excitations in the cluster operator (CCSD), but we highlight a recent preprint that has included two-electron, one-nucleus excitations in the cluster operator.[10] It is well known in single-component CC theory that CCSD is not sufficiently accurate for many applications and that connected triple-excitation contributions should be included in these cases.[2] Therefore, in this Letter, we derive and implement the perturbative multicomponent (T) correction for the first time. Multicomponent CCSD(T) is shown to accurately predict protonic affinities compared to experimental data and predict an FHF⁻ potential energy surface (PES) that agrees well with benchmark values. Additionally, absolute multicomponent CCSD(T) FHF⁻ energies are more accurate than multicomponent CCSD.

The starting point of the single-component (T) correction,[1] which we label $(T)_{ee}$ because it only includes electron-electron correlation contributions, is the single-component [T] correction (or CCSD+T(CCSD)),[11] which we label $[T]_{ee}$. The $[T]_{ee}$ correction is similar to the fourth-order triple-excitation energy term in single-component fourth-order Møller-Plesset perturbation theory (MP4)[12-13] but with converged double amplitudes from the single-component CCSD calculation. Assuming a canonical Hartree-Fock (HF) reference, the $[T]_{ee}$ and $(T)_{ee}$ corrections and corresponding triple-excitation amplitudes are shown diagrammatically in Figure 1. These diagrams are evaluated using the standard rules from many-body perturbation theory.[2] The $(T)_{ee}$ correction includes an additional diagram that is a fifth-order contribution in many-body perturbation theory,[1] but the additional diagram can also be derived by other means.[14] Algebraic expressions for these terms can be found in the existing literature.[2]

$$E_{[T]_{ee}} = \text{(diagram)}$$

$$E_{(T)_{ee}} = E_{[T]_{ee}} + \text{(diagram)}$$

$$\text{(amplitude diagram)} = \text{(amplitude diagram)}$$

**Figure 1:** Diagrams for the $[T]_{ee}$ and $(T)_{ee}$ corrections and single-component triple-excitation amplitudes.

In multicomponent CC theory, the terms that occur from purely electronic excitations are identical to the single-component case. Therefore, the only new terms that occur in the multicomponent (T) correction are purely nuclear-excitation terms and mixed electron-nuclear excitation terms. The purely nuclear terms are identical in structure to the purely electronic terms, but with nuclear triple-excitation amplitudes and two-particle nuclear integrals. For systems with only a single quantum nucleus, such as all calculations in this study, no purely nuclear-excitation terms occur. Because of this, the only new terms that need to be defined are the mixed electron-nuclear excitation terms. In this study, we derive a [T] and (T) perturbative correction from these mixed electronic-nuclear excitations, which is denoted as $[T]_{en}$ or $(T)_{en}$, respectively.

In this Letter, we assume a single nucleus is treated quantum mechanically, but the generalization to multiple quantum nuclei is straightforward. As a result, triple-excitation contributions from two-electron one-nucleus excitations need to be included while one electron, two-nuclei excitations do not. Using the multicomponent generalization of the single-component diagrammatic rules[4, 15] and assuming a multicomponent HF reference, the $[T]_{en}$ and $(T)_{en}$ corrections and two-electron, one-nucleus triple-excitation amplitudes are shown diagrammatically in Figure 2. In Figure 2, solid lines correspond to electronic hole or particle lines and dotted lines correspond to nuclear hole or particle lines. Comparing Figures 1 and 2, the similarities between single-component and multicomponent CC theory are immediately evident. We have derived spin-adapted algebraic equations from Figure 2 and implemented them into our existing open-source multicomponent CCSD code that is available from our GitHub repository.[16] In multicomponent CCSD(T), the (T) correction is equal to $(T) = (T)_{ee} + (T)_{en}$ with the [T] correction defined analogously. The $[T]_{en}$ correction is similar to the fourth-order triple-excitation contribution in multicomponent MP4,[15] but with converged double-excitation multicomponent CCSD amplitudes rather than first-order wave function contributions.

$$E_{[T]_{en}} = \text{(diagram)}$$

$$E_{(T)_{en}} = E_{[T]_{en}} + \text{(diagram)} + \text{(diagram)}$$

$$\text{(diagram)} = \text{(diagram)} + \text{(diagram)} + \text{(diagram)}$$

**Figure 2:** Diagrams for the $[T]_{en}$ and $(T)_{en}$ correction and mixed two-electron, one-nucleus triple-excitation amplitudes in multicomponent coupled-cluster theory.

We define the three terms in Figure 2 algebraically:

$$E_{[T]_{en}} = \frac{1}{4} \sum_{ijabIA} \left(t_{ijI}^{abA}\right)^2 * \varepsilon_{ijI}^{abA}, \qquad (1)$$

$$E_{(T)_{en}} = E_{[T]_{en}} - \sum_{ijabIA} t_i^a * t_{ijI}^{abA} * \langle jI|bA \rangle + \frac{1}{4} \sum_{ijabIA} t_I^A * t_{ijI}^{abA} * \langle ij||ab \rangle, \qquad (2)$$

and

$$\varepsilon_{ijI}^{abA} t_{ijI}^{abA} = -\hat{P}(ij) \sum_k t_{ik}^{ab} \langle kI|jA \rangle + \hat{P}(ab) \sum_c t_{ij}^{ac} \langle bI|cA \rangle - \hat{P}(ab) \sum_k t_{kI}^{aA} \langle kb||ij \rangle \qquad (3)$$

$$+ \hat{P}(ij) \sum_c t_{iI}^{cA} \langle ab||cj \rangle - \hat{P}(ab)P(ij) \sum_J t_{ij}^{aA} \langle jI|bJ \rangle + \hat{P}(ab)P(ij) \sum_B t_{iI}^{aB} \langle jA|bB \rangle.$$

In Eqs. 1-3 $t_i^a$, $t_I^A$, $t_{ij}^{ab}$ and $t_{iI}^{aA}$ are converged electronic single excitation; nuclear single excitation; two-electron excitation; and one-electron, one-nucleus excitation amplitudes; respectively. The operator $\hat{P}(pq)$ is equal to $\hat{P}(pq) = \hat{1} - \hat{P}_{pq}$, where $\hat{P}_{pq}$ permutes the indices $p$ and $q$. $\varepsilon_{ijI}^{abA} = \epsilon_i + \epsilon_j + \epsilon_I - \epsilon_a - \epsilon_b - \epsilon_A$, where $\epsilon_p$ is the molecular orbital energy of orbital $p$. The indices $i$, $j$, and $k$ label occupied electronic spin orbitals and $a$, $b$, and $c$ label virtual electronic spin orbitals. The convention for protonic orbitals is identical to the electronic orbitals but using upper-case letters. We adopt the convention that two-particle electron-proton integrals have a positive value.

The partitioning of the (T) correction into electron-electron correlation and electron-nuclear correlation contributions enables additional approximations to be made. In multicomponent methods, the nuclear properties are normally of primary interest, and it has been shown in multicomponent many-body methods that the electron-nuclear correlation has a greater effect on the nuclear properties than the electron-electron correlation.[15, 17] In most multicomponent calculations, the number of occupied nuclear orbitals is equal to 1 and the number of virtual nuclear orbitals is much smaller than the number of electronic virtual orbitals. Because of this, the computational bottleneck for a multicomponent many-body calculation where the electron-electron and electron-nuclear correlation are treated at the same level of theory is the electron-electron contributions. Treating the electron-electron correlation at a lower level of theory can greatly reduce the computational cost. Therefore, in this study, we also benchmark just the (T)en and [T]en corrections. Additionally, in systems that treat only a single nucleus quantum mechanically, the number of nuclear orbitals is fixed as the size of the system increases. Thus, the $(T)_{en}$ and $[T]_{en}$ corrections computationally scale $N^5$ with respect to system size but with an increased prefactor equal to the number of virtual nuclear orbitals compared to the purely electronic $N^5$ terms.

We have benchmarked the multicomponent $CCSD(T)_{en}$, $CCSD[T]_{en}$, and CCSD(T) methods by calculating the proton affinities of 12 small molecules with the most acidic hydrogen nucleus treated quantum mechanically and by calculating the energy of the FHF$^-$ molecule as a function of the F—F distance with the hydrogen nucleus treated quantum mechanically. Both are standard benchmarks for many-body multicomponent methods.[3]

The proton affinity of species A with only the transferring proton treated quantum mechanically is calculated as

$$\text{PA(A)} = \text{E}_\text{A} - \text{E}_{\text{AH}^+} + \frac{5}{2}RT, \tag{4}$$

where $\text{E}_\text{A}$ is calculated using single-component methodology and $\text{E}_{\text{AH}^+}$ is calculated with its multicomponent analogue. More detail about the derivation of Eq. 1 can be found in previous multicomponent studies.[3, 8] We briefly note that Eq. 1 assumes that the rotational energy of the reactants and products is identical and that the vibrational energy of the classical nuclei does not change upon protonation. These assumptions have previously been shown to be reasonable for multicomponent CCSD.[8]

To calculate $\text{E}_\text{A}$, single-component CCSD and CCSD(T) geometry optimizations were performed using CFOUR[18] and the aug-cc-pVDZ, aug-cc-pVTZ, and aug-cc-pVQZ electronic basis sets.[19-20] To determine $\text{E}_{\text{AH}^+}$, multicomponent CCSD, CCSD[T]$_{en}$, CCSD(T)$_{en}$, and CCSD(T) calculations were performed with the most acidic hydrogen nucleus of each molecule treated quantum mechanically. Calculations were performed with the aug-cc-pVDZ, aug-cc-pVTZ, and aug-cc-pVQZ electronic basis sets and a mixed electronic basis set where the quantum proton used the aug-cc-pVQZ electronic basis set and all other atoms used the aug-cc-pVTZ basis set. The latter basis set is labeled as the "mixed" electronic basis set in this Letter and has seen extensive use in previous multicomponent many-body calculations[7, 9, [15, 17] to lower the computational expense of the calculation while still including high angular-momentum basis functions that are thought to be necessary to correctly describe electron-nuclear correlation. The coordinates for the multicomponent calculations were obtained from a CCSD or CCSD(T) single-component geometry optimization of the AH$^+$ species using CFOUR with the same electronic basis set. An exception is that the mixed-basis multicomponent calculations used the aug-cc-pVTZ single-component geometry optimization result, which is consistent with the original multicomponent CC study.[7] Calculations were performed with the PB4D and PB4F protonic basis sets[21] centered at the location of the most acidic classical hydrogen atom in the single-component geometry optimization.

The energy of the FHF$^-$ molecule with the hydrogen nucleus treated quantum mechanically is calculated as a function of the F—F bond distance using the aug-cc-pVQZ electronic basis set and the PB4F protonic basis set. These results are compared to Fourier-grid Hamiltonian (FGH) calculations[22-23] with the single-point energy calculations performed at either the single-component CCSD or CCSD(T) levels of theory with the aug-cc-pVQZ electronic basis set. To be consistent with previous benchmarking,[7] the electronic basis functions are allowed to move with the hydrogen atom, which is a possible source of error,[24] but the FGH calculations are otherwise assumed to be nearly exact for a given electronic level of theory and electronic basis set. That is, the FGH calculations include all electron-nuclear correlation energy at the complete protonic basis-set limit.

The proton affinities for multicomponent CCSD are shown in Table 1. Our results with the mixed electronic and the PB4D protonic basis sets are identical to previous multicomponent CCSD results,[7-8] which validates our multicomponent CCSD code. The mean absolute error (MAE) and maximum absolute error (MaxAE) for the test set decrease as the size of the electronic basis set increases from aug-cc-pVDZ to aug-cc-pVTZ to aug-cc-pVQZ. Despite this decrease, the MAE for the aug-cc-pVQZ electronic basis set is large with values of 0.17 and 0.16 eV for the PB4D and PB4F protonic basis sets, respectively. The mixed basis set does not follow this trend. It has an MAE that is a factor of 4 lower than the MAE for the aug-cc-pVQZ basis set calculations for both the PB4D and PB4F protonic basis sets. Given the smooth behavior of the MAEs as the size of the electronic basis set is increased, it appears possible that part of the reason for the previous good performance of multicomponent CCSD is due to use of the mixed electronic basis set. We hypothesize that the larger basis set on the quantum proton lowers the basis set error for only a subset of the system and this lower energy contribution cancels the error in the electron-nuclear correlation energy in multicomponent CCSD. That is, multicomponent CCSD with the mixed electronic basis set appears to be a Pauling point[25] for proton affinities.

| Molecule | Experiment | PB4D | | | | PB4F | | | |
|---|---|---|---|---|---|---|---|---|---|
| | | aug-cc-pVDZ | aug-cc-pVTZ | aug-cc-pVQZ | mixed | aug-cc-pVDZ | aug-cc-pVTZ | aug-cc-pvQZ | mixed |
| $CN^-$ | 15.31 | -0.59 | -0.31 | -0.26 | -0.11 | -0.58 | -0.30 | -0.24 | -0.09 |
| $NO_2^-$ | 14.75 | -0.41 | -0.21 | -0.15 | 0.01 | -0.40 | -0.20 | -0.13 | 0.03 |
| $NH_3$ | 8.85 | -0.36 | -0.21 | -0.13 | -0.05 | -0.36 | -0.20 | -0.11 | -0.03 |
| $HCOO^-$ | 14.97 | -0.44 | -0.23 | -0.16 | -0.01 | -0.44 | -0.22 | -0.14 | 0.01 |
| $HO^-$ | 16.95 | -0.46 | -0.22 | -0.14 | 0.00 | -0.46 | -0.21 | -0.12 | 0.02 |
| $HS^-$ | 15.31 | -0.56 | -0.32 | -0.26 | -0.10 | -0.55 | -0.32 | -0.25 | -0.09 |
| $H_2O$ | 7.16 | -0.42 | -0.24 | -0.17 | -0.06 | -0.42 | -0.23 | -0.15 | -0.04 |
| $H_2S$ | 7.31 | -0.35 | -0.20 | -0.15 | 0.00 | -0.34 | -0.20 | -0.13 | 0.01 |
| CO | 6.16 | -0.41 | -0.22 | -0.18 | -0.04 | -0.41 | -0.21 | -0.16 | -0.03 |
| $N_2$ | 5.12 | -0.44 | -0.25 | -0.19 | -0.07 | -0.44 | -0.24 | -0.17 | -0.05 |
| $CO_2$ | 5.60 | -0.38 | -0.23 | -0.18 | -0.03 | -0.37 | -0.22 | -0.16 | -0.01 |
| $CH_2O$ | 7.39 | -0.36 | -0.21 | -0.13 | -0.05 | -0.36 | -0.20 | -0.11 | -0.03 |
| MAE | | 0.43 | 0.24 | 0.17 | 0.04 | 0.43 | 0.23 | 0.16 | 0.04 |
| MaxAE | | 0.59 | 0.32 | 0.26 | 0.11 | 0.58 | 0.32 | 0.25 | 0.09 |

**Table 1:** Multicomponent CCSD proton affinities errors relative to experiment, mean absolute error (MAE), and maximum absolute error (MaxAE) for different electronic and protonic basis sets. A negative value indicates that the calculated value is less than the experimental value. Experimental values are from references 26-30. All values are in eV.

Additional evidence for this claim is seen in Tables 2, 3, and 4, which show the proton affinity errors for the multicomponent CCSD[T]$_{en}$, CCSD(T)$_{en}$, and CCSD(T) methods, respectively. Like multicomponent CCSD, the MAE and MaxAE for these methods monotonically decrease as the electronic basis set is increased from aug-cc-pVDZ to aug-cc-pVTZ to aug-cc-pVQZ.

However, except for multicomponent CCSD(T) with the PB4D protonic basis set, the CCSD[T]$_{en}$, CCSD(T)$_{en}$, and CCSD(T) methods have larger MAEs for calculations with the mixed electronic basis set than calculations with the aug-cc-pVQZ electronic basis set, which starkly contrasts to the results from the multicomponent CCSD calculations. Even in the case of multicomponent CCSD(T) with the PB4D protonic basis set, the lower MAE for the mixed electronic basis set compared to the aug-cc-pVQZ electronic basis set is likely due to a fortuitous cancelation of error. Because the aug-cc-pVTZ proton affinities for multicomponent CCSD(T) with the PB4D protonic basis set are all smaller than the experimental value, increasing the basis set size solely on the quantum proton lowers the energy by an amount sufficient to give better agreement with the experimental values.

| Molecule | Experiment | PB4D | | | | PB4F | | | |
|---|---|---|---|---|---|---|---|---|---|
| | | aug-cc-pVDZ | aug-cc-pVTZ | aug-cc-pVQZ | mixed | aug-cc-pVDZ | aug-cc-pVTZ | aug-cc-pvQZ | mixed |
| $CN^-$ | 15.31 | -0.52 | -0.18 | -0.08 | 0.06 | -0.51 | -0.16 | -0.02 | 0.12 |
| $NO_2^-$ | 14.75 | -0.32 | -0.08 | 0.03 | 0.19 | -0.32 | -0.05 | 0.08 | 0.24 |
| $NH_3$ | 8.85 | -0.29 | -0.10 | 0.02 | 0.10 | -0.29 | -0.09 | 0.06 | 0.13 |
| $HCOO^-$ | 14.97 | -0.29 | -0.10 | 0.02 | 0.10 | -0.29 | -0.09 | 0.06 | 0.13 |
| $HO^-$ | 16.95 | -0.36 | -0.10 | 0.01 | 0.16 | -0.35 | -0.07 | 0.06 | 0.21 |
| $HS^-$ | 15.31 | -0.49 | -0.18 | -0.06 | 0.10 | -0.48 | -0.16 | 0.00 | 0.15 |
| $H_2O$ | 7.16 | -0.36 | -0.14 | -0.03 | 0.08 | -0.35 | -0.11 | 0.01 | 0.12 |
| $H_2S$ | 7.31 | -0.28 | -0.08 | 0.03 | 0.18 | -0.28 | -0.06 | 0.08 | 0.22 |
| CO | 6.16 | -0.35 | -0.11 | -0.02 | 0.11 | -0.34 | -0.08 | 0.03 | 0.16 |
| $N_2$ | 5.12 | -0.37 | -0.14 | -0.04 | 0.08 | -0.37 | -0.11 | 0.01 | 0.13 |
| $CO_2$ | 5.60 | -0.30 | -0.11 | -0.03 | 0.11 | -0.30 | -0.09 | 0.02 | 0.17 |
| $CH_2O$ | 7.39 | -0.39 | -0.09 | 0.03 | 0.16 | -0.38 | -0.07 | 0.09 | 0.22 |
| MAE | | 0.36 | 0.12 | 0.03 | 0.12 | 0.36 | 0.10 | 0.04 | 0.17 |
| MaxAE | | 0.52 | 0.18 | 0.08 | 0.19 | 0.51 | 0.16 | 0.09 | 0.24 |

**Table 2:** Multicomponent CCSD[T]$_{en}$ proton affinities errors relative to experiment, mean absolute error (MAE), and maximum absolute error (MaxAE) for different electronic and protonic basis sets. Experimental values are from references 26-30. A negative value indicates that the calculated value is less than the experimental value. All values are in eV.

| Molecule | Experiment | PB4D | | | | PB4F | | | |
|---|---|---|---|---|---|---|---|---|---|
| | | aug-cc-pVDZ | aug-cc-pVTZ | aug-cc-pVQZ | mixed | aug-cc-pVDZ | aug-cc-pVTZ | aug-cc-pvQZ | mixed |
| $CN^-$ | 15.31 | -0.52 | -0.20 | -0.11 | 0.03 | -0.52 | -0.18 | -0.06 | 0.08 |
| $NO_2^-$ | 14.75 | -0.32 | -0.09 | 0.01 | 0.17 | -0.32 | -0.07 | 0.05 | 0.21 |
| $NH_3$ | 8.85 | -0.30 | -0.22 | 0.00 | 0.07 | -0.30 | -0.10 | 0.03 | 0.10 |
| $HCOO^-$ | 14.97 | -0.30 | -0.22 | 0.00 | 0.07 | -0.30 | -0.10 | 0.03 | 0.10 |
| $HO^-$ | 16.95 | -0.36 | -0.11 | -0.01 | 0.14 | -0.35 | -0.09 | 0.04 | 0.19 |
| $HS^-$ | 15.31 | -0.49 | -0.21 | -0.10 | 0.06 | -0.49 | -0.19 | -0.05 | 0.11 |
| $H_2O$ | 7.16 | -0.36 | -0.15 | -0.05 | 0.06 | -0.36 | -0.13 | -0.02 | 0.09 |
| $H_2S$ | 7.31 | -0.29 | -0.10 | 0.00 | 0.14 | -0.29 | -0.08 | 0.04 | 0.18 |
| CO | 6.16 | -0.36 | -0.12 | -0.05 | 0.08 | -0.35 | -0.11 | -0.01 | 0.12 |
| $N_2$ | 5.12 | -0.38 | -0.15 | -0.07 | 0.05 | -0.37 | -0.13 | -0.02 | 0.09 |
| $CO_2$ | 5.60 | -0.31 | -0.13 | -0.05 | 0.09 | -0.30 | -0.10 | -0.01 | 0.14 |
| $CH_2O$ | 7.39 | -0.39 | -0.11 | 0.01 | 0.14 | -0.38 | -0.08 | 0.06 | 0.19 |
| MAE | | 0.36 | 0.15 | 0.04 | 0.09 | 0.36 | 0.11 | 0.03 | 0.13 |
| MaxAE | | 0.52 | 0.22 | 0.11 | 0.17 | 0.52 | 0.19 | 0.06 | 0.21 |

**Table 3:** Multicomponent CCSD(T)$_{en}$ proton affinities errors relative to experiment, mean absolute error (MAE), and maximum absolute error (MaxAE) for different electronic and protonic basis sets. Experimental values are from references 26-30. A negative value indicates that the calculated value is less than the experimental value. All values are in eV.

| Molecule | Experiment | PB4D | | | | PB4F | | | |
|---|---|---|---|---|---|---|---|---|---|
| | | aug-cc-pVDZ | aug-cc-pVTZ | aug-cc-pVQZ | mixed | aug-cc-pVDZ | aug-cc-pVTZ | aug-cc-pvQZ | mixed |
| $CN^-$ | 15.31 | -0.56 | -0.25 | -0.16 | -0.01 | -0.56 | -0.23 | -0.11 | 0.04 |
| $NO_2^-$ | 14.75 | -0.43 | -0.17 | -0.08 | 0.09 | -0.42 | -0.15 | -0.05 | 0.13 |
| $NH_3$ | 8.85 | -0.34 | -0.16 | -0.05 | 0.03 | -0.34 | -0.14 | -0.02 | 0.06 |
| $HCOO^-$ | 14.97 | -0.46 | -0.19 | -0.09 | 0.06 | -0.45 | -0.17 | -0.05 | 0.10 |
| $HO^-$ | 16.95 | -0.50 | -0.21 | -0.10 | 0.04 | -0.49 | -0.19 | -0.05 | 0.09 |
| $HS^-$ | 15.31 | -0.57 | -0.25 | -0.15 | 0.02 | -0.56 | -0.24 | -0.11 | 0.06 |
| $H_2O$ | 7.16 | -0.40 | -0.18 | -0.09 | 0.02 | -0.40 | -0.16 | -0.05 | 0.06 |
| $H_2S$ | 7.31 | -0.33 | -0.12 | -0.03 | 0.12 | -0.33 | -0.11 | 0.01 | 0.16 |
| CO | 6.16 | -0.36 | -0.13 | -0.05 | 0.08 | -0.35 | -0.11 | -0.01 | 0.12 |
| $N_2$ | 5.12 | -0.37 | -0.15 | -0.07 | 0.05 | -0.37 | -0.13 | -0.03 | 0.09 |
| $CO_2$ | 5.60 | -0.37 | -0.17 | -0.10 | 0.05 | -0.36 | -0.15 | -0.06 | 0.09 |
| $CH_2O$ | 7.39 | -0.34 | -0.16 | -0.05 | 0.03 | -0.34 | -0.14 | -0.02 | 0.06 |
| MAE | | 0.42 | 0.18 | 0.08 | 0.05 | 0.41 | 0.16 | 0.05 | 0.09 |
| MaxAE | | 0.57 | 0.25 | 0.16 | 0.12 | 0.56 | 0.24 | 0.11 | 0.16 |

**Table 4:** Multicomponent CCSD(T) proton affinities errors relative to experiment, mean absolute error (MAE), and maximum absolute error (MaxAE) for different electronic and protonic basis sets. Experimental values are from references 26-30. A negative value indicates that the calculated value is less than the experimental value. All values are in eV.

Given the results in Tables 1-4, we assume that the aug-cc-pVQZ electronic and PB4F protonic basis set results are the best representation of the relative accuracy of the different multicomponent CC methods. From this subset of results, it appears that connected triple-excitation contributions are essential for quantitative accuracy as multicomponent CCSD has an MAE over three times higher than the new methods introduced in this study. Multicomponent CCSD[T]$_{en}$ and CCSD(T)$_{en}$ and CCSD(T) are all of similar accuracy, yielding a maximum difference of 0.02 eV among the MAEs. Similar to previous studies,[15, 17] this finding shows that for properties that depend principally on NQEs of the quantum proton, the electron-electron correlation can often be treated at a lower level of theory than the electron-nuclear correlation with only small changes to proton affinities.

As was discussed in the original multicomponent CCSD study,[7] single-component CCSD(T) with the aug-cc-pVQZ electronic basis set performs well for the calculation of proton affinities with an MAE of 0.03 when using the same test set of 12 molecules. However, these single-component calculations require the calculation of a Hessian to include the zero-point energy, while the multicomponent formalism includes the zero-point energy from a single-point energy calculation. Additionally, many of the potential applications of multicomponent methods, such as the computation of vibrationally averaged PESs,[31-32] have no single-component analogue. Thus, the introduction of more accurate multicomponent methods, such as the multicomponent CC methods of this study, is therefore essential for such future applications.

The energy of the FHF$^-$ molecule as a function of the F—F distance with the hydrogen nucleus treated quantum mechanically for different multicomponent CC methods is shown in Figures 3 and 4. As the multicomponent CCSD, CCSD[T]$_{en}$, and CCSD(T)$_{en}$ methods include no purely electronic connected triple-excitation contributions, they are benchmarked relative to FGH calculations at the single-component CCSD level of theory. The multicomponent CCSD(T) method is benchmarked relative to FGH calculations at the single-component CCSD(T) level of theory.

The inclusion of NQEs results in a minimum on the PES with a longer F—F distance for all methods compared to the single-component minimum. All the multicomponent methods overestimate this outward shift relative to the FGH method, but multicomponent CCSD is the most accurate as it overestimates the shift by approximately 0.005 Å compared to errors of 0.015, 0.01, or 0.015 Å for the multicomponent CCSD[T]$_{en}$, CCSD(T)$_{en}$, or CCSD(T) methods, respectively. This too large shift in the F—F distance compared to the FGH methods upon inclusion of triple-excitation contributions to the electron-nuclear correlation energy was also seen in the previous study of multicomponent MP4,[15] but the cause of this error remains unknown. We hypothesize it may in part be due to the treatment of the electronic basis set in the FGH calculations,[24] as was mentioned previously. This theory may warrant future investigation.

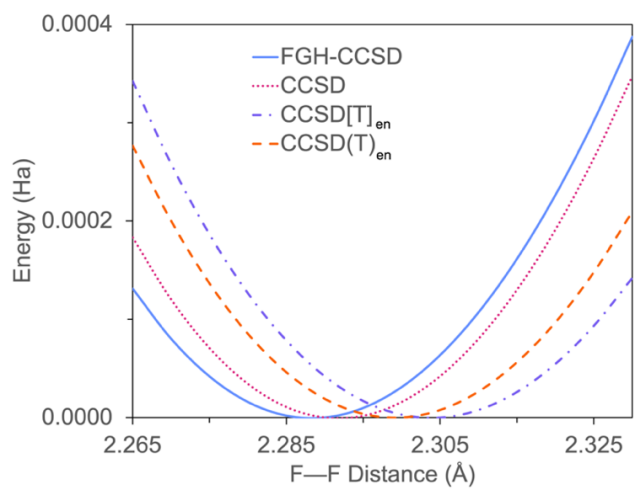

**Figure 3**: Energy of the FHF$^-$ molecule as a function of the F—F distance for the FGH method using single-component CCSD (solid blue), multicomponent CCSD (dotted magenta), multicomponent CCSD[T]$_{en}$ (dashed-dotted purple), and multicomponent CCSD(T)$_{en}$ (dashed orange). For each method, the minimum energy F—F distance is set to 0.0 Ha.

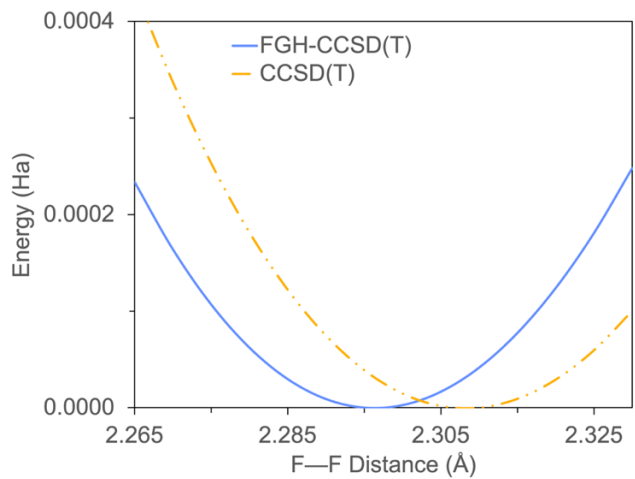

**Figure 4:** Energy of the FHF$^-$ molecule as a function of the F—F distance for the FGH method using single-component CCSD(T) (solid blue) and multicomponent CCSD(T) (dashed-double dotted yellow). For each method, the minimum energy F—F distance is set to 0.0 Ha.

As an additional measure of the accuracy of the multicomponent CC methods, we have computed the absolute error of the multicomponent CCSD, CCSD[T]$_{en}$, CCSD(T)$_{en}$, and CCSD(T) methods for the FHF$^-$ molecule as a function of the F—F distance relative to the FGH method as shown in Figure 5. In Figure 5, the error of multicomponent CCSD(T) is relative to FGH calculations with single-component CCSD(T), while all other methods are relative to FGH calculations with single-component CCSD. From Figure 3 demonstrates that including electronic-nuclear triple excitations in multicomponent CC calculations results in much better absolute energies for this system. Multicomponent CCSD[T]$_{en}$ and CCSD(T)$_{en}$ have errors that are reduced by factors of 3 and 10, respectively, compared to the multicomponent CCSD method. Multicomponent CCSD(T) has a smaller absolute error than all other multicomponent CC methods in this study. The CCSD>CCSD[T]$_{en}$>CCSD(T)$_{en}$>CCSD(T) ordering of absolute errors for multicomponent CC is what would be intuitively expected based on the typical relative accuracy of single-component CC methods.[2] This observation provides further evidence that the relative accuracy of multicomponent methods is similar to the relative accuracy of their single-component analogues as was hypothesized in the recent multicomponent MP4 study.[15]

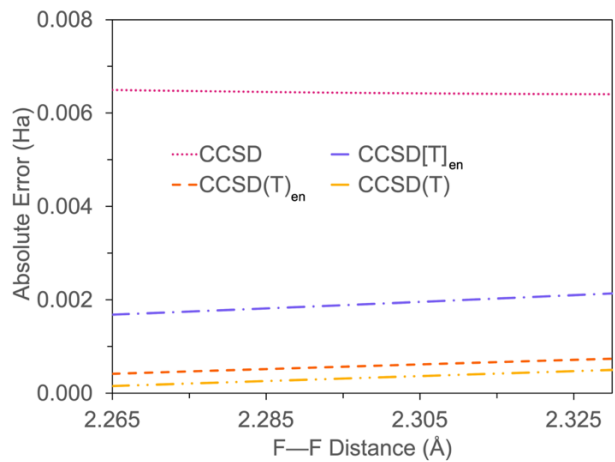

**Figure 5:** Absolute error of the multicomponent CCSD (dotted magenta), multicomponent CCSD[T]$_{en}$ (dashed-dotted purple), multicomponent CCSD(T)$_{en}$ (dashed orange), and multicomponent CCSD(T) (dashed-double dotted yellow) for calculations on the FHF- molecules as a function of the F—F distance. For multicomponent CCSD(T), FGH with single-component CCSD(T) was used as the reference, while all other multicomponent methods used FGH with single-component CCSD as the reference.

Finally, we discuss the computational timings of the multicomponent CC methods in this study. All performance-critical code in our program is written in Cython. Our multicomponent CCSD implementation is a multicomponent generalization of the single-component CCSD algorithm of Scuseria and coworkers[33] and should be relatively well optimized. Our single-component (T)$_{ee}$ correction is based on the implementation in PySCF,[34] which uses a *abcijk* batch implementation,[35] and so it should also be relatively well optimized. At present, our implementation of the (T)$_{en}$ correction is not computed over batches and likely could be further optimized. Even without any optimization, the (T)$_{en}$ correction is not the computational bottleneck for any of the systems in this study. As was previously discussed, this is because the (T)$_{en}$ correction scales $N^5$ with respect to system size calculation with a single quantum nucleus because the number of occupied and virtual nuclear orbitals is constant. A representative selection of timings from this study is shown in Table 5. All calculations were performed on a single Intel Xeon Gold 6252 2.10 GHz core. For all calculations using the aug-cc-pVDZ electronic basis set, the (T)$_{en}$ correction takes more wall time than a single iteration of the purely electronic amplitude equations. For calculations using the aug-cc-pVQZ electronic basis set, the (T)$_{en}$ correction takes significantly less time than the solution of a single iteration of the purely electronic amplitude equations. This empirically demonstrates that the (T)$_{en}$ correction can be included in most realistic multicomponent CC calculations without being the computational bottleneck, because this Letter shows that electronic basis sets of at least quadruple-zeta quality are needed for accurate protonic properties.

| System | Electronic Basis | Electronic Basis Functions | Occupied Electronic Basis Functions | Amplitude Eq. | Elec. Amplitude Eq. | $(T)_{ee}$ | $(T)_{en}$ |
|---|---|---|---|---|---|---|---|
| $H_2O$ | aug-cc-pVDZ | 37 | 5 | 0.4 | 0.3 | 2.4 | 1.3 |
|  | aug-cc-pVTZ | 92 |  | 20.6 | 19.5 | 51.4 | 20.4 |
|  | aug-cc-pVQZ | 172 |  | 359.3 | 353.4 | 429.9 | 138.3 |
| $H_3O^+$ | aug-cc-pVDZ | 46 | 5 | 1.0 | 0.9 | 5.1 | 2.5 |
|  | aug-cc-pVTZ | 115 |  | 74.1 | 71.9 | 45.0 | 112.5 |
|  | aug-cc-pVQZ | 218 |  | 967.2 | 1046.6 | 1178.4 | 365.4 |
| $HCO^+$ | aug-cc-pVDZ | 55 | 7 | 4.3 | 3.7 | 10.0 | 7.4 |
|  | aug-cc-pVTZ | 115 |  | 68.0 | 64.7 | 143.5 | 69.6 |
|  | aug-cc-pVQZ | 206 |  | 1183.8 | 1156.5 | 1407.0 | 520.6 |
| HCOOH | aug-cc-pVDZ | 83 | 12 | 58.6 | 54.3 | 106.7 | 75.2 |
|  | aug-cc-pVTZ | 184 |  | 1357.4 | 1306.9 | 2323.7 | 801.3 |
|  | aug-cc-pVQZ | 332 |  | 37692.7 | 37119.1 | 27948.8 | 5514.1 |

**Table 5:** Representative wall times for multicomponent CCSD(T) calculations in this study. All times in seconds. "Amplitude equation" is the time for one iteration of the multicomponent CCSD amplitude equations. "Electronic amplitude equation" is the time for one iteration of the purely electronic amplitude equation in a multicomponent CCSD calculation. All calculations are performed using the PB4F protonic basis set.

In conclusion, we have derived and implemented the (T) correction to multicomponent CCSD and used it to compute the proton affinities of 12 small molecules and the energy of the FHF⁻ molecule at different F—F distances. It was shown that previous good results for multicomponent CCSD for proton affinities may be in part due to the use of a mixed electronic basis set and a cancellation of errors. An approximation to the full (T) correction, called $(T)_{en}$, is introduced that only includes triple-excitation contributions from mixed electronic-protonic excitations. The multicomponent CCSD$(T)_{en}$ method is similar in accuracy to the multicomponent CCSD(T) method for the prediction of the nuclear properties in this study but has a significantly reduced computational cost. Given the wide use of the single-component CCSD(T) method, we expect the multicomponent CCSD$(T)_{en}$ and CCSD(T) methods to see extensive use for benchmarking other multicomponent methods and in application studies of small molecules.

**Author Information**
Corresponding Author
*E-mail: brorsenk@missouri.edu

**Acknowledgements**
K.R.B. thanks the University of Missouri for startup funding.

**Notes**
The authors declare no other competing financial interest.

**References**
(1) Raghavachari, K.; Trucks, G. W.; Pople, J. A.; Head-Gordon, M. A Fifth-Order Perturbation Comparison of Electron Correlation Theories. *Chem. Phys. Lett.* **1989,** *157*, 479-483.
(2) Shavitt, I.; Bartlett, R. J. *Many-Body Methods in Chemistry and Physics: MBPT and Coupled-Cluster Theory*. Cambridge University Press: 2009.
(3) Pavošević, F.; Culpitt, T.; Hammes-Schiffer, S. Multicomponent Quantum Chemistry: Integrating Electronic and Nuclear Quantum Effects Via the Nuclear–Electronic Orbital Method. *Chem. Rev.* **2020,** *120*, 4222-4253.
(4) Nakai, H.; Sodeyama, K. Many-Body Effects in Nonadiabatic Molecular Theory for Simultaneous Determination of Nuclear and Electronic Wave Functions: Ab Initio NOMO/MBPT and CC Methods. *J. Chem. Phys.* **2003,** *118*, 1119-1127.
(5) Ellis, B. H.; Aggarwal, S.; Chakraborty, A. Development of the Multicomponent Coupled-Cluster Theory for Investigation of Multiexcitonic Interactions. *J. Chem. Theory Comput.* **2015,** *12*, 188-200.
(6) Ellis, B. H.; Aggarwal, S.; Chakraborty, A. Development of the Multicomponent Coupled-Cluster Theory for Investigation of Multiexcitonic Interactions. *J. Chem. Theory Comput.* **2016,** *12*, 188-200.
(7) Pavošević, F.; Culpitt, T.; Hammes-Schiffer, S. Multicomponent Coupled Cluster Singles and Doubles Theory within the Nuclear-Electronic Orbital Framework. *J. Chem. Theory Comput.* **2018,** *15*, 338-347.
(8) Pavošević, F.; Hammes-Schiffer, S. Multicomponent Coupled Cluster Singles and Doubles and Brueckner Doubles Methods: Proton Densities and Energies. *J. Chem. Phys.* **2019,** *151*, 074104.
(9) Pavošević, F.; Hammes-Schiffer, S. Multicomponent Equation-of-Motion Coupled Cluster Singles and Doubles: Theory and Calculation of Excitation Energies for Positronium Hydride. *J. Chem. Phys.* **2019,** *150*, 161102.
(10) Pavošević, F.; Hammes-Schiffer, S. Triple Electron-Electron-Proton Excitations and Second-Order Approximations in Nuclear-Electronic Orbital Coupled Cluster Methods. *arXiv:2206.13616v1* **2022**.
(11) Urban, M.; Noga, J.; Cole, S. J.; Bartlett, R. J. Towards a Full CCSDT Model for Electron Correlation. *J. Chem. Phys.* **1985,** *83*, 4041-4046.
(12) Krishnan, R.; Pople, J. A. Approximate Fourth-Order Perturbation Theory of the Electron Correlation Energy. *Int. J. Quantum Chem.* **1978,** *14*, 91-100.
(13) Krishnan, R.; Frisch, M.; Pople, J. Contribution of Triple Substitutions to the Electron Correlation Energy in Fourth Order Perturbation Theory. *J. Chem. Phys.* **1980,** *72*, 4244-4245.
(14) Stanton, J. F. Why CCSD (T) Works: A Different Perspective. *Chem. Phys. Lett.* **1997,** *281*, 130-134.
(15) Fajen, O. J.; Brorsen, K. R. Multicomponent MP4 and the Inclusion of Triple Excitations in Multicomponent Many-Body Methods. *J. Chem. Phys.* **2021,** *155*, 234108.
(16) https://github.com/brorsenk/mc_coupled_cluster.
(17) Fajen, O. J.; Brorsen, K. R. Separation of Electron–Electron and Electron–Proton Correlation in Multicomponent Orbital-Optimized Perturbation Theory. *J. Chem. Phys.* **2020,** *152*, 194107.
(18) Matthews, D. A.; Cheng, L.; Harding, M. E.; Lipparini, F.; Stopkowicz, S.; Jagau, T. C.; Szalay, P. G.; Gauss, J.; Stanton, J. F. Coupled-Cluster Techniques for Computational Chemistry: The CFOUR Program Package. *J. Chem. Phys.* **2020,** *152*, 214108.
(19) Dunning Jr, T. H. Gaussian Basis Sets for Use in Correlated Molecular Calculations. I. The Atoms Boron through Neon and Hydrogen. *J. Chem. Phys.* **1989,** *90*, 1007-1023.

(20) Kendall, R. A.; Dunning Jr, T. H.; Harrison, R. J. Electron Affinities of the First-Row Atoms Revisited. Systematic Basis Sets and Wave Functions. *J. Chem. Phys.* **1992,** *96*, 6796-6806.
(21) Yu, Q.; Pavošević, F.; Hammes-Schiffer, S. Development of Nuclear Basis Sets for Multicomponent Quantum Chemistry Methods. *J. Chem. Phys.* **2020,** *152*, 244123.
(22) Marston, C. C.; Balint-Kurti, G. G. The Fourier Grid Hamiltonian Method for Bound State Eigenvalues and Eigenfunctions. *J. Chem. Phys.* **1989,** *91*, 3571-3576.
(23) Webb, S. P.; Hammes-Schiffer, S. Fourier Grid Hamiltonian Multiconfigurational Self-Consistent-Field: A Method to Calculate Multidimensional Hydrogen Vibrational Wavefunctions. *J. Chem. Phys.* **2000,** *113*, 5214-5227.
(24) Alaal, N.; Brorsen, K. R. Multicomponent Heat-Bath Configuration Interaction with the Perturbative Correction for the Calculation of Protonic Excited States. *J. Chem. Phys.* **2021,** *155*, 234107.
(25) Lowdin, P.-O. Some Comments on the Present Situation of Quantum Chemistry in View of the Discussions at the Dubrovnik Workshop on the Electronic Structure of Molecules. *Pure Appl. Chem.* **1989,** *61*, 2185-2194.
(26) Cumming, J. B.; Kebarle, P. Summary of Gas Phase Acidity Measurements Involving Acids AH. Entropy Changes in Proton Transfer Reactions Involving Negative Ions. Bond Dissociation Energies D (A—H) and Electron Affinities EA (a). *Can. J. Chem.* **1978,** *56*, 1-9.
(27) Ervin, K. M.; Ho, J.; Lineberger, W. C. Ultraviolet Photoelectron Spectrum of Nitrite Anion. *J. Phys. Chem.* **1988,** *92*, 5405-5412.
(28) Graul, S. T.; Schnute, M. E.; Squires, R. R. Gas-Phase Acidities of Carboxylic Acids and Alcohols from Collision-Induced Dissociation of Dimer Cluster Ions. *Int. J. Mass Spectrom. Ion Processes* **1990,** *96*, 181-198.
(29) Hunter, E. P.; Lias, S. G. Evaluated Gas Phase Basicities and Proton Affinities of Molecules: An Update. *J. Phys. Chem. Ref. Data* **1998,** *27*, 413-656.
(30) Jolly, W. L. *Modern Inorganic Chemistry*. Mcgraw Hill International Edition: 2017.
(31) Xu, X.; Yang, Y. Constrained Nuclear-Electronic Orbital Density Functional Theory: Energy Surfaces with Nuclear Quantum Effects. *J. Chem. Phys.* **2020,** *152*, 084107.
(32) Xu, X.; Yang, Y. Full-Quantum Descriptions of Molecular Systems from Constrained Nuclear–Electronic Orbital Density Functional Theory. *J. Chem. Phys.* **2020,** *153*, 074106.
(33) Scuseria, G. E.; Janssen, C. L.; Schaefer Iii, H. F. An Efficient Reformulation of the Closed-Shell Coupled Cluster Single and Double Excitation (CCSD) Equations. *J. Chem. Phys.* **1988,** *89*, 7382-7387.
(34) Sun, Q.; Zhang, X.; Banerjee, S.; Bao, P.; Barbry, M.; Blunt, N. S.; Bogdanov, N. A.; Booth, G. H.; Chen, J.; Cui, Z.-H. Recent Developments in the PySCF Program Package. *J. Chem. Phys.* **2020,** *153*, 024109.
(35) Rendell, A. P.; Lee, T. J.; Komornicki, A.; Wilson, S. Evaluation of the Contribution from Triply Excited Intermediates to the Fourth-Order Perturbation Theory Energy on Intel Distributed Memory Supercomputers. *Theor. Chim. Acta* **1993,** *84*, 271-287.